\documentstyle[preprint,prl,aps,epsfig]{revtex}
\def\beq{\begin{equation}}
\def\eeq{\end{equation}}
\def\beqa{\begin{eqnarray}}
\def\eeqa{\end{eqnarray}}

\def\zb{\beta}
\def\lsim{\mathrel{\raise.3ex\hbox{$<$\kern-.75em\lower1ex\hbox{$\sim$}}} }
\def\gsim{\mathrel{\raise.3ex\hbox{$>$\kern-.75em\lower1ex\hbox{$\sim$}}} }

\begin{document}
\draft
\preprint{{\vbox{\hbox{NCU-HEP-k003} \hbox{Dec 2001}\hbox{rev. Mar 2002} }}}
%@twocolumnfalse\endcsname

%\addtocounter{page}{-1}

\title {Observing CP Violating MSSM Higgs Bosons at Hadron Colliders ? }
\author{\bf Abdesselam Arhrib} 
\address{Faculty of Sciences and Techniques, \\Department of Mathematices, 
B.P. 416, Tangier, Morocco\\
\& UFR High Energy Physics, Physics Department , \\Faculty of Sciences,
B.P. 1014, Rabat, Morocco
\\E-mail: arhrib@fstt.ac.ma}

\author{\bf Dilip Kumar Ghosh}
\address{Department of Physics, \\National Taiwan University, Taipei, TAIWAN 10617
\\E-mail: dghosh@phys.ntu.edu.tw}

\author{\bf Otto C. W. Kong}
\address{Department of Physics, \\National Central University, Chung-li, TAIWAN 32054
\\E-mail: otto@phy.ncu.edu.tw}
%\date{\today}
\maketitle

{%\tighten
\begin{abstract}
We report on the possibility of observing Higgs sector CP violation of the minimal
supersymmetric standard model at a hadron machine. The CP phase dependent
cross-sections for the $VH_i$ associated production processes are given for
the Fermilab Tevatron and CERN LHC. Substantial production cross-sections
for channels of all three Higgs bosons simultaneously are shown to be 
possible, giving a direct indication of the CP violation. The observability of the 
Higgs signals are discussed.
\end{abstract}
}
% ? \pacs{12.15.Lk, 12.38.Bx, 12.60.Fr,  14.80.Cp }
%\vskip0pc]
%\vskip2pc]

%\narrowtext
\newpage

%\section*{Introduction}
The search for the Higgs bosons, scalar particles from the electroweak symmetry
breaking multiplet(s), in the Standard Model (SM)  and beyond is a major goal of 
present and  future colliders. One or more light Higgs boson within relatively
easy reach of the upgraded Tevatron or CERN LHC is particularly favorable
to the most popular extension of the SM, namely the minimal 
supersymmetric standard model (MSSM). Hence, it is very important
that we study all possible scenarios under the framework in careful
details. Here in this letter, we make an effort in the direction focusing
on the scenario with radiatively induced Higgs sector CP violation\cite{rcp,rscp}.

In the recent years, the so-called CP violating MSSM has became a subject 
of many phenomenological studies. A major part of the latter focus on Higgs 
physics, especially with application to the LEP machine at CERN 
(see, for example, Ref.\cite{CEPW} and references therein). 
Implications for a $e^+e^-$ machine is quite well studied. In particular,
Ref.\cite{AA} has pushed the analysis to the prospective Next Linear Collider.
Nevertheless, after the closing of the LEP machine and before another lepton machine 
is commissioned, we have no choice but to focus on the not as clean environment
of the hadron colliders. Hence, it is the time for detailed careful studies of
the topic at hadron machines. Some steps in the direction have been taken.
More notable ones include works on aspects of the production phase in 
Refs.\cite{DM,CL} restricting to the gluon fusion mechanism\cite{gun},
as well as analysis of the subsequent decays of the Higgs bosons 
produced\cite{decay,decay2}. A complete analysis of the collider signature from 
production to decays with the inclusion of signal-background studies is of course 
the final goal. However, the topic is a complicated one, and may have to be taken 
one step at a time.\footnote{During the preparation of the present letter, a study
combining the production and decay processes comes up\cite{CHL}.
The latter reference is also restricted to production through gluon fusion
and is concentrated on the lightest Higgs boson.}
The present letter reports the first step by the present authors in the direction.
We aim here at presenting explicit production cross-sections for all the three neutral 
Higgs bosons through the Higgstrahlung processes from quark-quark collisions, 
{\it i.e.}, $q\,\bar{q}\longrightarrow Z^{{\scriptscriptstyle 0}*}\longrightarrow Z^{\scriptscriptstyle 0}\, H_i$ and
$q\,\bar{q}' \longrightarrow W^{{\scriptscriptstyle \pm}*}\longrightarrow W^{\scriptscriptstyle \pm}\, H_i$. 
The gluon fusion process is generally expected to have the largest cross-section 
but suffer from very large background\cite{tvtH}.  Hence, other production channels 
may also prove to be useful in probing the CP violation in MSSM Higgs physics. 
It has been emphasized, in Ref.\cite{AA} for example, that due to the 
absence of an $A\,V\,V$ coupling, the simultaneous observation of all three
$V^*\longrightarrow V\, H_i$ channels will be a strong indication of
scalar-pseudoscalar mixings, and hence Higgs sector CP violation, in the MSSM.
We want to emphasize that this would be a qualitative result, pretty much
independent of the details of the exact cross-sections themselves and the
explicit determination of which region of parameter space the model lives in.
Seeing all the three Higgs channels basically says that the MSSM without the
Higgs sector CP violation is not right. One can always go on to models with a
richer Higgs spectrum. The CP violating scenario studied here would, however,
preferred by many as the viable alternative. 
We present here production cross-sections of the processes at both the Tevatron
and LHC, as our first probe towards the possibility of signals for the CP violation
at the two machines. The observability of such Higgs signals is a deeply involved,
but obviously very important, question. While a detailed quantitative study is beyond
our present report, we will try our best to address the issues involved qualitatively,
drawing lessons and comparison from related results in the literature. We would 
particularly like to draw attention to some plausibly important aspects beyond 
what has been studied in the literature. 

It should be noted that CP violating phases in MSSM are stringently constrained
by their contribution to electric dipole moments (EDMs). The topic has been
studied extensively\cite{edm}. It suffices here to emphasize that the EDM
constraints can be by-passed, for instance, by effectively decoupling the sparticles
of the first family and/or cancellations among the various contributions. 
Such constraints are not explicitly imposed in the present study. The rationale being
that the Higgs sector CP violation involves
flavor dependent parameters only of the third family and the complex phase
combination $\Phi_{\!\mbox{\tiny CP}}={\rm arg}(A\mu)$. This certainly leaves much
room for getting around the EDM constraints, by tuning the other parameters and 
phases for instance.

%\section*{2. Scalar-Pseudoscalar mixing in the MSSM}
The tree-level Higgs potential of the MSSM conserves CP. This ensures that
the three neutral Higgs mass eigenstates can be divided into the CP-even
$h^{\scriptscriptstyle 0}$ and $H^{\scriptscriptstyle 0}$ and CP-odd 
$A^{\scriptscriptstyle 0}$. The 1-loop effective potential, however, may violate CP. 
When this happens, three Higgs mass eigenstates with no definite CP parity would
be resulted. The Higgs bosons are denoted by 
$H^{\scriptscriptstyle 0}_{\!\scriptscriptstyle 1}$,
$H^{\scriptscriptstyle 0}_{\!\scriptscriptstyle 2}$, 
and $H^{\scriptscriptstyle 0}_{\!\scriptscriptstyle 3}$ (in ascending order of mass).
It has been shown that the CP violation may be generated by complex phases residing 
in the $\mu$ term and the soft SUSY breaking parameters $A_t$ (and $A_b$). 
These phases generate contributions to the off-diagonal block  
${\cal M}^2_{\mbox{\tiny SP}}$
in neutral Higgs mass-squared matrix ${\cal M}^2_{ij}$ mixing the scalar and 
pseudoscalar fields. These may be given approximately by\cite{PW}
\begin{eqnarray}
{\cal M}^2_{\mbox{\tiny SP}} &\approx& {\cal O}
\left({m^4_t \; |\mu||A_t| \over v^2 \; 32\pi^2 \; M^2_{\mbox{\tiny SUSY}}}\right) \; \sin\!\Phi_{\!\mbox{\tiny CP}} 
\nonumber \\
&& \times \left[6,\;{|A_t|^2\over M^2_{\mbox{\tiny SUSY}}}, \;
{|\mu|^2\over \tan\!\beta \; M^2_{\mbox{\tiny SUSY}}}, \;
{\sin\!2\Phi_{\!\mbox{\tiny CP}}  \; |A_t| \; |\mu| 
\over \sin\!\Phi_{\!\mbox{\tiny CP}} \; M^2_{\mbox{\tiny SUSY}}}\right] \;,
\end{eqnarray}
where $\Phi_{\!\mbox{\tiny CP}}={\rm arg}(A_t\mu)$. Here, we have only displayed 
the contributions from the top squarks ($\tilde t_{\scriptscriptstyle 1,2}$)
which are dominant for small $\tan\!\beta$\cite{Ltb}. Sizeable scalar-pseudoscalar 
mixing is possible for large $|\mu|$ and $|A_t| $ ($> M_{\mbox{\tiny SUSY}}$).

%\section{Calculation and Results}
In our numerical Higgs mass computation, we follow Ref.\cite{CEPW} and use the 
public code  available at \cite{cph}.  This involves one-loop effective potential 
with large two-loop non-logarithmic corrections induced by one-loop threshold 
effects on the top and bottom quark Yukawa couplings included. We are interested in
regions of parameter space where all three Higgs boson of masses are relatively close
to each other, say all smaller than $200\,\mbox{GeV}$. Otherwise, the model would be 
close to the decoupling limit where the radiative CP effect on the Higgs sector is 
known to be unimportant\cite{rcp,CEPW}. We are also restricting to relatively small
$\tan\!\beta$ value, with demonstrated  substantial scalar-pseudoscalar mixings. 
For simplicity, we take nonzero a common phase for $A_t$ and $A_b$
as the sole source of  $\Phi_{\!\mbox{\tiny CP}}$.

At the patron level, to the leading order (LO), the cross-section for a $VH_i$ 
associated production process is given by
\begin{eqnarray} \label{cs}
\hat{\sigma}_{\!\mbox{\tiny LO}}(q\bar{q^\prime} \to V H_i )
= \frac{G_{\scriptscriptstyle F}^2 \, M_{\scriptscriptstyle V}^4}{288 \,\pi\, Q^2}(v_q^2 +a_q^2) \;
\frac{\lambda(m_{\scriptscriptstyle V}^2,m_{\scriptscriptstyle {H_i}}^2,Q^2)
+12 \,m_{\scriptscriptstyle V}^2/Q^2}{(1-m_{\scriptscriptstyle V}^2/Q^2)^2} 
\sqrt{\lambda(m_{\scriptscriptstyle V}^2,m_{\scriptscriptstyle {H_i}}^2,Q^2)}\ C_i^2
\end{eqnarray}
where $\lambda(x,y,z)=(1-x/z-y/z)^2-4 x y/z^2$ and $v_q=-a_q=\sqrt{2}$ for 
$V=W^{\scriptscriptstyle \pm}$ and 
$v_q= 2 \, I_3^q - 4 \,e_{\!q} \, \sin\!^2\!\theta_{\!\scriptscriptstyle W}$, \ $ a_q=2 I_3^q$ for 
$V=Z^{\scriptscriptstyle 0}$ ($q^\prime =q$); while $C_i$ is the $VVH_i$ coupling 
renormalized to ${g\, M_{\!\scriptscriptstyle Z}\over\cos\!\theta_{\!\scriptscriptstyle W}}$.
A crucial point here is that the three Higgs bosons mix through an ``orthogonal" 
matrix, leading to a sum rule for the $C_i$ couplings\cite{sr}. Namely,
\beq \label{sr}
C_1^2 + C_2^2 + C_3^2=1 \, .
\eeq
The sum rule is well appreciated among researchers on the subject. It guarantees 
that at least one of the three production cross-sections is not suppressed. This 
feature is much exploited  in Higgs discovery studies. The sum rule also suggests 
that all the three cross-sections can be simultaneously substantial for some 
particular set of the relevant SUSY parameters. The latter feature plays a central 
role in our present analysis. Explicit presentations of the variations of the $C_i$'s,
or $VVH_i$ couplings, for the CP violating case of interest here have been given 
in many of the earlier works\cite{sr2}. Hence, we skip explicit numerical 
presentation here.

We convolute the admissible patron sub-process cross-section given above with the 
CTEQ4L patron distribution functions. QCD corrections are known to give an 
enhancement of about 40\% for the Tevatron (Run II) and 30\% for the 
LHC\cite{qcd}. Other SUSY corrections are generally small\cite{sqcd}. In the 
latter reference, it is shown to be less than $1.5\%$,  being smaller for large 
squark/gluino masses. In the explicit plots given in the figures, we scaled the 
LO cross-sections calculated with the corresponding enhancement  $K$ 
($={\sigma_{\!\mbox{\tiny NLO}} / \sigma_{\!\mbox{\tiny LO}} }$) factor,
to give a better indication of the next-leading order (NLO) results. The $K$ factor
is taken simply to be $1.4$ and $1.3$ for the Tevatron and LHC, respectively. This
should be good enough for the present purpose. Readers interested in further details
on the tiny variations in the exact $K$ factor value along with the changes of the model
parameters are referred to Ref.\cite{sqcd}.

We illustrate our results with two representative set of chosen input parameters.
Parameter Set~A is chosen in accordance with the  benchmark scenario (CPX) 
introduced in Ref.\cite{bmk} aimed at maximizing the CP violating effects.
The CPX scenario is as follows:
\begin{eqnarray}
& & \widetilde{M}_Q=\widetilde{M}_t=\widetilde{M}_b= 1\,{\mbox{TeV}} \ , \ 
\mu = 4 \,\mbox{TeV}  \ ,\; |A_t|=|A_b|= 2\, \mbox{TeV} \ , \nonumber \\
& &\ |m_{\tilde{g}}|= 1\, \mbox{TeV} \quad  {\mbox{and}} \quad 
\ |m_{\widetilde{B}}|= |m_{\widetilde{W}}|= 0.3\,\mbox{TeV} \;,
\end{eqnarray}
added to which we fixed  
\[ \tan\!\zb=6 \]
for Set~A inputs.  Parameter Set~B is similar, with little variations, namely
\begin{eqnarray}
& & \widetilde{M}_Q=\widetilde{M}_t=\widetilde{M}_b=1\,{\mbox{TeV}}\,,
\; \mu = 2\,\mbox{TeV}\ , \; |A_t|=|A_b|= 2\,\mbox{TeV}\ ,\nonumber \\
& & \ |m_{\tilde{g}}|= 1\,\mbox{TeV} \quad {\mbox{and}} \quad 
\ |m_{\widetilde{B}}|= |m_{\widetilde{W}}|= 0.3\,\mbox{TeV}\;, \nonumber \\
& & \qquad \mbox{with} \qquad \qquad \tan\!\zb= 15\;.
\end{eqnarray}
The charged Higgs mass is taken as the control parameter on the scale of the 
Higgs masses. We show only results for two choices of  the charged Higgs mass,
at $150\,\mbox{GeV}$ and $200\,\mbox{GeV}$. The $150\,\mbox{GeV}$
gives $H_1$ mass very close to the known bound from LEP. The  
$200\,\mbox{GeV}$ case gives a relatively high mass scale value getting close to,
while still staying away from, the decoupling limit. Hence, the two choices
roughly envelope the range of interest. The production cross-sections are plotted 
as a function of $\Phi_{\!\mbox{\tiny CP}}$, which comes here from a common 
phase of  $A_t$ and $A_b$. Explicit plots of the masses are also given for easy 
cross-reference. Fig.~1 and 2 are results for the Tevatron, based on Set~A and 
Set~B inputs, respectively, while Fig.~3 and 4 give the corresponding results for 
the LHC. The figures do illustrate the existence of the exciting possibility we go 
after, namely,  having substantial production cross-sections simultaneously for all 
the three $VH_i$ channels. This should not come as a surprise. Naively, the sum 
rule [{\it cf.} Eq.(\ref{sr})] suggests there might be a ``democratic" limit where 
the three channels share the overall coupling equally. In that situation, each 
$H_i$ would have a production cross-section at the same order of magnitude as
that of the SM Higgs of the same mass, only suppressed by a small factor. 
Our results do indicate explicitly that the scenario could be more or less
achieved for some optimal $\Phi_{\!\mbox{\tiny CP}}$ value. 

Let us also briefly comment on the basic features of the plots. The general
features of the dependence of the masses and gauge couplings [represented by
the $C_i$'s in Eq.(\ref{cs}) above] of the three (physical) Higgs bosons
upon the CP phase $\Phi_{\!\mbox{\tiny CP}}$ through the stop mixing parameter 
$|X_t|=|A_t - \mu\,\cot\!\zb|$ have been studied by various groups 
(see Ref.\cite{CEPW} and references therein). In each of the top panels of  
Figs. 1-4, our explicitly illustrated Higgs masses agree well with previous 
studies.  A particularly note-worthy aspect is that the (one loop corrected) 
$H_1$ mass increases with  $|X_t|$ till reaching its maximum at 
$|X_t|/M_{\tiny{\mbox{squark}}} \simeq 2.45$, and drops with further increase
in $|X_t|$. Here within each panel, the latter is tuned with 
$\Phi_{\!\mbox{\tiny CP}}\simeq {\rm arg}(A_t)$. In the plot
of Fig.~1(a), for example, $H_1$ mass is maximum at 
$\Phi_{\!\mbox{\tiny CP}}\simeq 80^o$. The large effect of the CP phase enhancing 
stop mixing here promotes scalar-pseudoscalar mixings, hence suppresses $C_1$.
This, together with the  $H_1$ mass enhancement, gives a minimum for the $VH_1$ 
production cross-sections at $\Phi_{\!\mbox{\tiny CP}}\simeq 88^o$, as
shown in the plots right below [Fig.~1(b) and 1(c)]. 
$\!\!\!$\footnote{
More explicitly, from the relation $C_i=\cos\! \beta \, O_{1i} + \sin\!\beta \, O_{2i}$,
we have $C_1 \approx \sin\!\beta \, O_{21} $  for large $\tan\!\beta$.  
$C_1$  is then dominated by $O_{21}$ which flips sign.
That is why we have a dip in the plot. 
}
The particle actually assumes
the character of the pseudoscalar around this point. The situation is almost
completely reversed for $H_2$, which simply corresponds to the pseudoscalar
at the vanishing $\Phi_{\!\mbox{\tiny CP}}$ limit. In this case, $H_3$ is not
much affected by  variation in  $\Phi_{\!\mbox{\tiny CP}}$ except when it 
assumes the character of the pseudoscalar  at the other CP conserving limit
of $\Phi_{\!\mbox{\tiny CP}}= 180^o$. With larger mass splitting between
$H_2,H_3$ and the lightest Higgs $H_1$ as given by the second case in
Fig.~1 [plots 1(d), 1(e), and 1(f)], we are getting closer to the decoupling limit. 
The $H_1$ then behaves like the SM Higgs. It not much affected by   
$\Phi_{\!\mbox{\tiny CP}}$ variation, which now mainly describes mixing between
the two heavier states.  With the same set of inputs, the features in Fig.~3 are
more or less the same as in Fig.~1. Fig.~2 and 4 are results from a different 
input set of parameters (Set~B).  The set of input parameters is not very 
different from the previous case though, as we are strongly confined by our
special interest in large $\Phi_{\!\mbox{\tiny CP}}$ induced mass mixings.
The point where $VH_1$  cross-sections are strongly suppressed while 
$VH_2$  cross-sections enhanced in Fig.~2 ({\it cf.} Fig.~1) is now shifted to 
$\Phi_{\!\mbox{\tiny CP}}\simeq 110^o$. In all the cases, roughly at the
central values of $\Phi_{\!\mbox{\tiny CP}}$  between two of the dips of the
three $VH_i$  cross-section  curves, one gets the solutions for substantial 
cross-sections for all three Higgs channels simultaneously.

%Comments
One of our major result is that in the most favorable situation, there is a chance 
that all three Higgs boson could be simultaneously produced with around or above 
$0.01\,\mbox{pb}$ cross-sections at the Tevatron. The mass region of interest to 
us is in fact happens to be just well covered by the machine. This suggests the 
exciting possibility of seeing Higgs sector CP violation, assuming MSSM. 
At the LHC, the cross-sections could all go simultaneously to around or above 
$0.1\,\mbox{pb}$. The more or less optimal case correspond to results illustrated
in the sets of left panels in the figures, with masses for all three $H_i$ below
$150\,\mbox{GeV}$. The sets of right panels in the figures, however, illustrate
roughly the cases of limiting capacity. Here, $H_2$ and $H_3$ are a bit heavier.
They are around $180/190\,\mbox{GeV}$. As we will discuss below, this might be
really pushing the limit on signal observability. However, it is our opinion that 
a careful and detailed analysis is required to set the definite mass reach. The 
latter may get somewhat close to the situation illustrated in these panels.
This is especially true in the case of LHC. 

One certainly should not be too optimistic about the scenario. We put the 
question mark in the title of this letter because there are good reasons to be
cautious. Assuming that the MSSM really falls into such a parameter space
region where the three CP violating Higgs boson can all be produced with
substantial cross-sections at the Tevatron or LHC, identifying the signals
may be daunting task. In the spirit of our present analysis, one can claim that 
the CP violating Higgs bosons are unambiguously observed only after we have 
successfully identified all the three $H_i$'s produced through their decays.
Otherwise, one may have to rely on details on cross-section measurements
and further inputs from other possible SUSY signals (if at all available) to pin 
down the Higgs signals as coming from CP violating MSSM. We emphasize
again that identifying three Higgs mass eigenstates produced through
the $V\,H_i$ channels in itself establishes Higgs sector CP violation,
{\it i.e.} assuming MSSM. With anything less than that, it is going to be
an extremely involved task to confirm that the Higgs boson(s) observed
is/are more than generic MSSM Higgs boson(s).

Observing ``intermediate mass" neutral Higgs bosons are notoriously difficult 
(see, for example, Ref.\cite{gun}).  The ``gold-plated" 4-lepton decay mode
$H\to Z^0Z^0\to l^+l^+l^-l^-$\cite{GKW} has good branching ratio only for heavier 
Higgs bosons, while the 2$\,\gamma$ mode $H\to\gamma\gamma$  has steeply 
increasing background\cite{GKW}. Associated productions, the presently considered 
$VH$\cite{GKW,KKS} , $\bar{t}tH$\cite{ttH}, or even $\tilde{t}\tilde{t}H$\cite{sttH}
have been advocated as viable alternatives to the gluon fusion process to be 
the focus of Higgs hunting exactly for that reason.  A recent paper\cite{BBS} gives 
detailed discussions on the various associated production processes focusing on the 
effects of large stop mixing. The latter, while closer to our present study, is still
limited to the CP-conserving MSSM. This is also the case for  most of the previous 
studies\cite{orr}. Such studies, while of some use in giving a rough idea on the
observability of the Higgs signals under discussion, certainly cannot take the
place of the necessary specific studies, as along the line recently started by
Refs.\cite{decay,decay2}. However, results from the latter references are not
enough to help us to reach any definite conclusion. 

Ref.\cite{decay2} in particular gives interesting results on partial branching 
fractions of the various decays for the three $H_i$'s separately. It is a big step 
in the direction. What is still missing though are signal versus background 
analyses. The reference gives no definite mass reach numbers for the $H_i$'s. One 
would like to have the definite reach of the hadron machines in terms of the 
mass control parameter ({\it i.e.} charged Higgs mass) as a function of 
$\Phi_{\!\mbox{\tiny CP}}$ reflecting the range of simultaneous 
observability of all three Higgs signals. This ambitious task is beyond the present 
short letter. Mass reach numbers for MSSM Higgs bosons are not widely
available, even for the general case without Higgs sector CP violation. 
We can only present below discussions based on available information 
in the literature. The discussion aims at giving some idea on what might be 
expected. Hence, it may have some speculative element. We will quote some
mass reach number for the SM, or SM-like lightest MSSM, Higgs. These
numbers are of course {\it not} directly applicable to our scenario. However,
the SM case is much better studied. We have pointed out above that the
coupling sum rule [{\it cf.} Eq.(\ref{sr})] close to the ``democratic" limit suggests
a scenario in which the three $H_i$'s kind of share equally the role of the SM
Higgs. Each $H_i$ then behaves like ``a third of" the SM Higgs at the same mass.
If one take the latter statement seriously while assuming all the other
SUSY particles are heavier, we have a situation where the SM Higgs numbers
do provide a useful guideline.

The case for the Tevatron may be quite marginal but we consider it worth the 
effort to check it in details, focusing on both the  $H\to b\bar{b}$ and $H\to WW^{(*)}$
decays, or including even more decay channels. Decay branching fraction results
from Ref.\cite{decay2} do confirm the $H\to b\bar{b}$ as the dominating
channels at least up to masses of  $150\,\mbox{GeV}$. So why should we consider the 
$H\to WW^{(*)}$ and other channels? 

Current searches planned for Tevatron Run II actually concentrate on the $VH$ 
associated production process with decays $H\to b\bar{b}$\cite{SMW} for Higgs mass
in the lower intermediate range, taking the extra advantage that the leptonic decays of 
$W^\pm$ or $Z^0$ can be used as triggers to suppress background. SM Higgs is 
expected to be observable through the channel, however, only for mass up to 
$130\,\mbox{GeV}$ or slightly above\cite{KM}. For Higgs mass in the upper  intermediate 
range, $135-180\,\mbox{GeV}$, the planned focus is rather on production through gluon 
fusion with subsequent decay $H\to WW^{(*)}$\cite{HZ}. The reference claims that the mass 
reach for a SM like Higgs, with an integrated luminosity  of $30\,\mbox{fb}^{-1}$,  
would be up to  $190\,\mbox{GeV}$. 
$\!\!$\footnote{We should add that the  $\bar{t}tH$ with $H\to b\bar{b}$ has also been 
advocated as a discovery mode at Tevatron\cite{G6}.
}
A reason behind is the strong background for the $b\bar{b}$ signal. In our scenario, 
at least for masses of around or above  $150\,\mbox{GeV}$, $H_i\to WW^{(*)}$ 
could be sizeable for all three Higgs bosons\cite{decay2}. On the other hand,  $b\bar{b}$ 
coupling(s) would be suppressed for the heavier Higgs states. Obviously, we cannot 
rely on the $H\to b\bar{b}$ channels to see all the three $H_i$'s then. Nevertheless, 
unlike the SM case, using gluon fusion does not help confirming the CP violating 
setting we are interested in here. In this regard, a previous study on trilepton Higgs 
signal\cite{3l} is particularly relevant. We will very likely have to rely on the 
$WH \to WWW^{(*)} \to 3\,l$ decay to identify at least one or two of the $VH_i$ 
channels. With an integrated luminosity  of $100\,\mbox{fb}^{-1}$, Ref.\cite{3l}
claims a limiting $3\,\sigma$ reach for the SM Higgs in mass range
$140-175\,\mbox{GeV}$, with suggestions on further gains to be achieved. We
conclude that a combined study of the $VH$ production processes with decay 
channels and signal-background analyses specifically for the CP violating MSSM 
scenario will have to take the trilepton signal as one of the major focus.

The situation looks much better at the LHC.  Studies for SM-like Higgs shown that, 
for the $VH$ associated production under consideration, the $H\to b\bar{b}$ channel 
should have reasonable efficiency in identifying the signal\cite{km}, at least in the 
lower  intermediate mass range. $H\to\gamma\gamma$ would also be 
useful\cite{KKS,orr}. Again, in the upper intermediate mass range, even for a 
SM-like Higgs one will have to go back to 
$H\to WW^{(*)}$ (or $H\to Z^0Z^0\to l^+l^+l^-l^-$). In addition, we emphasize 
again that the $H_i b\bar{b}$ couplings are very unlike to be simultaneously 
unsuppressed. Hence, the $WH \to WWW^{(*)} \to 3\,l$ type trilepton signals are 
definitely useful for probing the CP violating model. Ref.\cite{3l} claims a 
$5\sigma$ discovery reach for the mass range 
$140-180 \,(125-200)\,\mbox{GeV}$ with $30\, (100)\,\mbox{fb}^{-1}$ for a SM 
Higgs. One may naively expect the signal for each $H_i$ to be weaken by a third 
or so in the optimal case of ``democratically shared" couplings (equal $C_i$'s). 
That sounds quite encouraging.

It should be noted that, in general, possible decays of $H_2$ and $H_3$ through $H_1$ 
itself  may compete with the $H_i\to WW^{(*)}$ channels and complicates analyses 
of the latter. However, in the region of parameter space of interest here, such decays 
are unlikely to be important and hence not taken into consideration here.
Finally, we should mention that decays into superparticles are likely to dominate
if their masses put them within kinematic threshold of the Higgs decays. This is 
very unlikely for the scenario we are interested in here, as one can easily see 
from the illustrative parameter input Set~A and B given above, hence not discussed.

Perhaps we should also mention a related production mechanism, the weak
boson fusion channels (see Ref.\cite{wbf} and references therein). Similar
to the associated productions, the processes exploit the $VVH$ couplings.
It is advocated as a possible Higgs discovery channel at LHC, and a useful
tool to determine the CP nature of the Higgs boson involved. From the 
present perspective, it will be interesting to explore the possibility of
simultaneous observation of all three $VV\to H_i$ channels as a
probe for the Higgs sector CP violation. 

In summary, we illustrate in this letter explicit results on the production
cross-sections of the three $VH_i$ channels at the Tevatron and the LHC. 
Our results indicate that a simultaneous observation of the three channels
may be a possibility, though could be quite marginal at the Tevatron. In
the best scenario, MSSM with radiative Higgs sector CP violation could
give rise to cross-sections of the order or above $0.01\,\mbox{pb}$ and
$0.1\,\mbox{pb}$,  for the Tevatron and the LHC, respectively. Detail
signal-background analyses exploring various decay modes are called
for.  Nevertheless, we hope that the above discussions have convinced the
readers that there are good reasons to be optimistic. 
In particular, we point out that the $WH_i \to WWW^{(*)} \to 3\,l$ 
decays are going to be useful. Assuming MSSM, the simultaneous 
observations of the three  $VH_i$ channels will confirm the  radiative 
Higgs sector CP violation scenario.

\section*{Acknowledgement}
We thank K.~Cheung, A.~Datta, S.~Raychaudhuri, S.~Sarkar, and
M.~Spira for helpful discussions, and K.~Cheung
for comments on a draft version of the manuscript.
D.~Ghosh, and O.~Kong are supported by the National Science
Council under the grants  NSC 90-2811-M-002-054,
and NSC 90-2112-M-008-051, respectively. D.~Ghosh also has support
from the Ministry of Education Academic Excellence Project 89-N-FA01-1-4-3.
We are also benefited from activities under the SUSY sub-program of
the Particle Physics program, National Center for Theoretical Sciences.
O.~Kong would particularly like to thank colleagues at NCTS and
Institute of Physics, Academia Sinica, especially S.~C. ~Lee and H.-N.~Li
for support.

%%%%%%%%%%%%%%%%%%%%%%%%%%%%%%%%%%%%%%%

\eject

\begin{figure}[h]
\vspace*{-2.5in}
\centerline{\epsfig{file=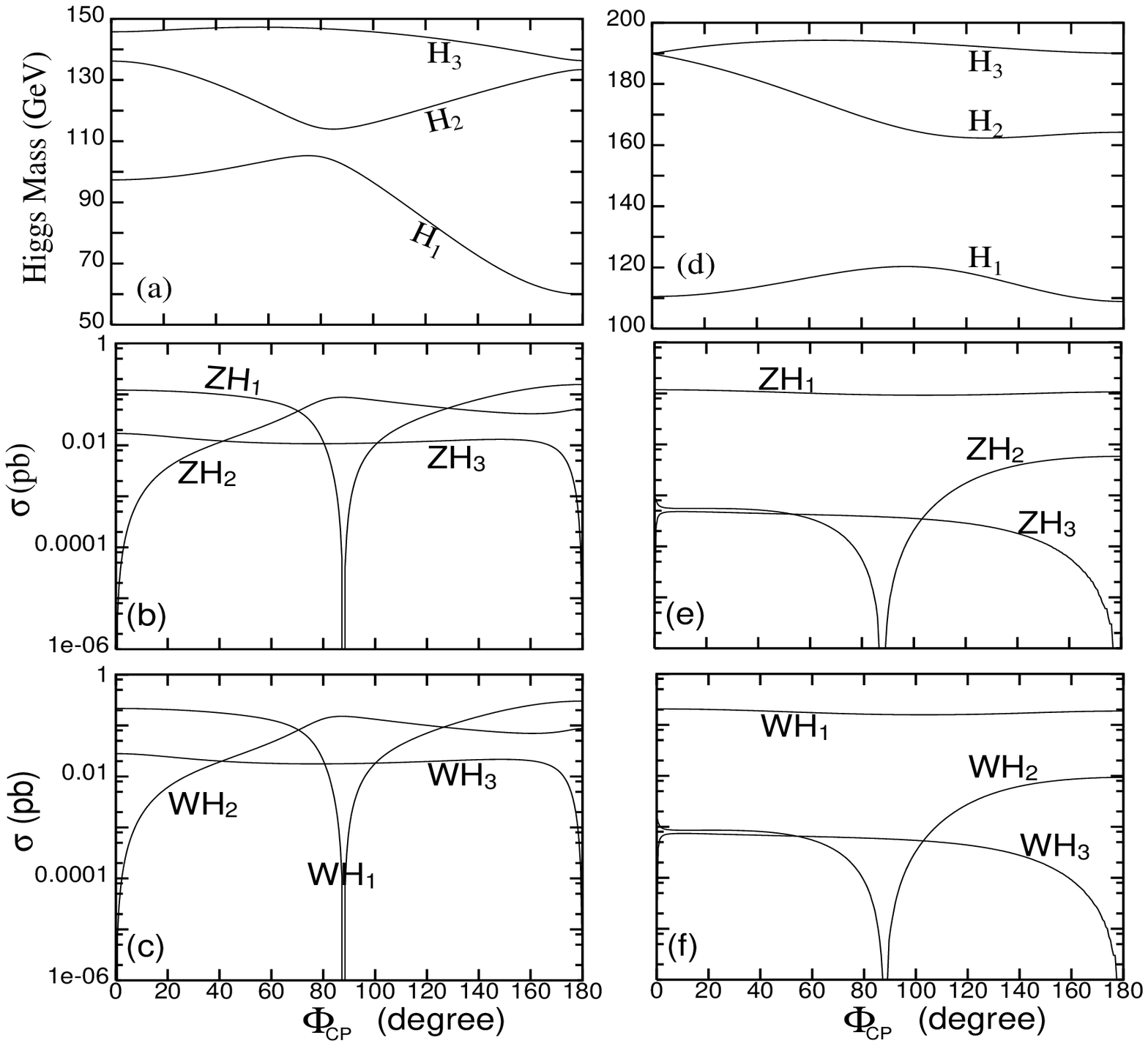,width=\linewidth}}
\caption{\small Plots  (a) and (d) show the variation of three Higgs masses 
             with the CP violating phase $\Phi_{\!\mbox{\tiny CP}}$. Plots  (b), (e) and 
             (c) ,(f) show the variation of $ZH_i$ and $WH_i$, $(i =1,2,3)$ production 
             cs at Tevatron Run II energy with the phase $\Phi_{\!\mbox{\tiny CP}}$. 
             The left panel and right panel correspond to the charged 
             Higgs mass of 150 GeV and 200 GeV
             respectively. Other MSSM parameter is fixed to Set~A.} 
\label{fig_1}
\end{figure}
\newpage
\begin{figure}[h]
\vspace*{-2.5in}
\centerline{\epsfig{file=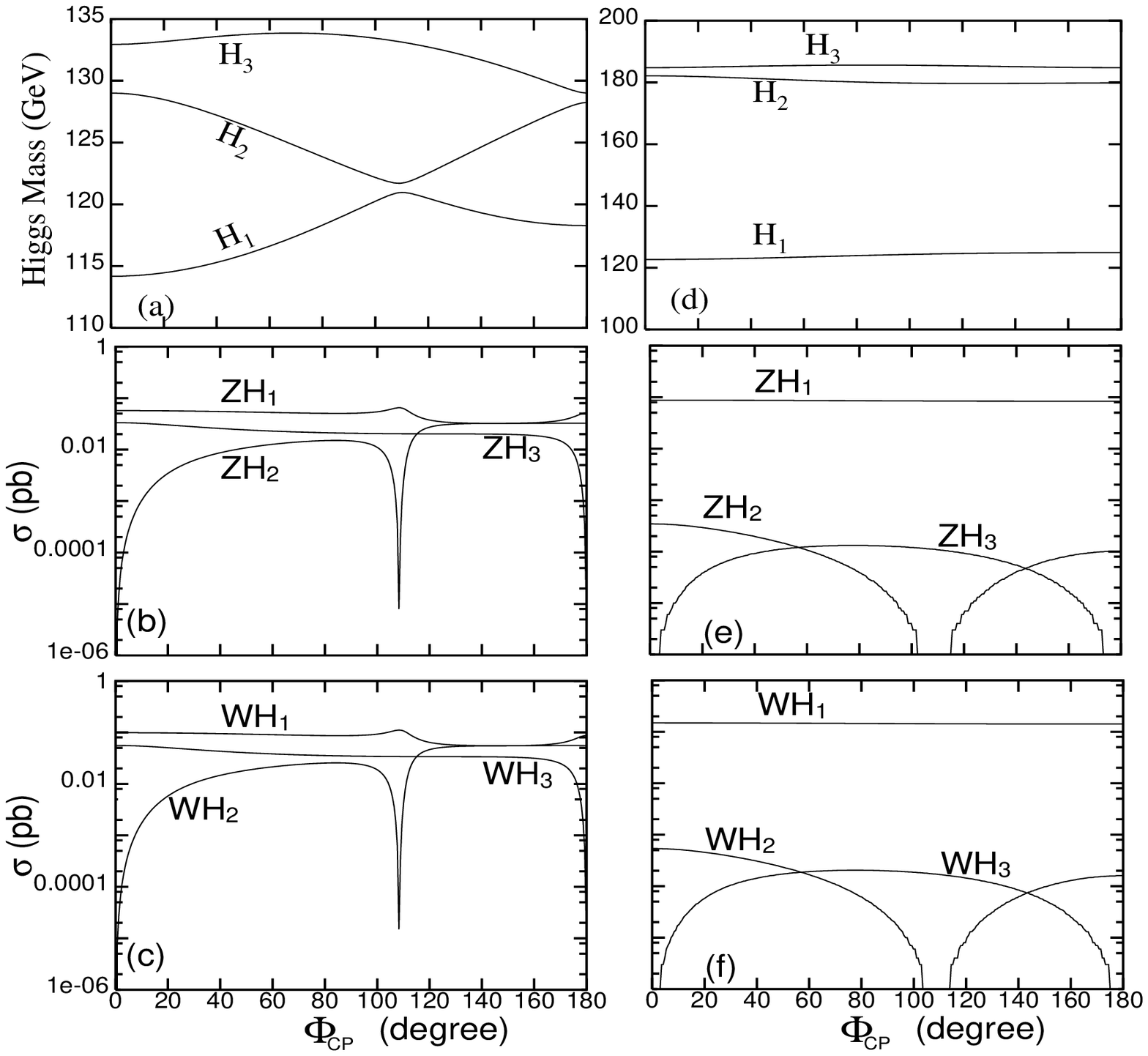,width=\linewidth}}
\caption{\small Plots  (a) and (d) show the variation of three Higgs masses 
             with the CP violating phase $\Phi_{\!\mbox{\tiny CP}}$. Plots (b), (e) and 
             (c) ,(f) show the variation of $ZH_i$ and $WH_i$, $(i =1,2,3)$  production 
            cs at Tevatron Run II energy with the phase $\Phi_{\!\mbox{\tiny CP}}$. 
             The left panel and right panel correspond to the charged Higgs
             mass of 150 GeV and 200 GeV respectively. Other MSSM parameter is 
             fixed to Set~B.} 
\label{fig_2}
\end{figure}
\eject
%\newpage
\begin{figure}[h]
\vspace*{-2.5in}
\centerline{\epsfig{file=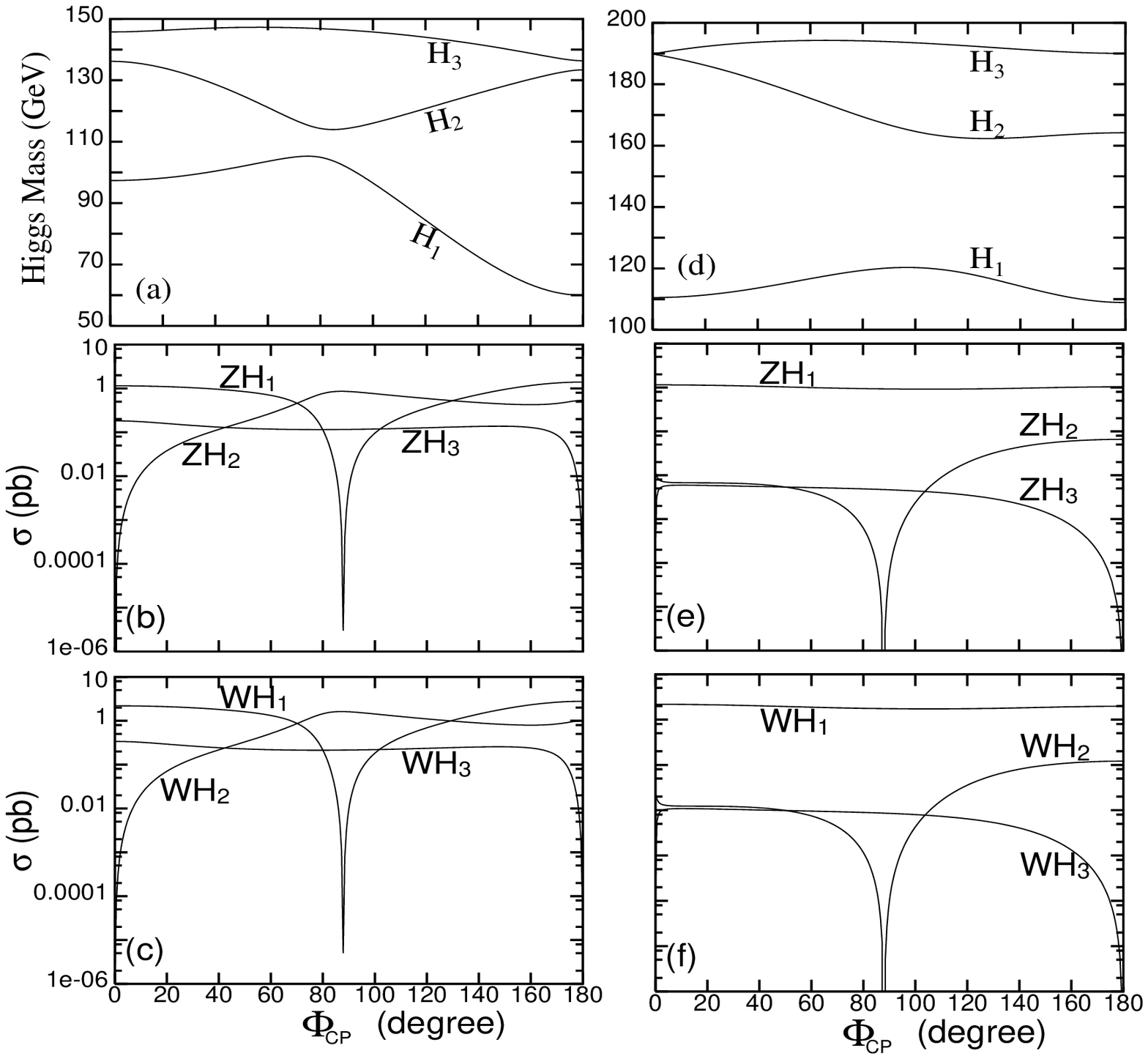,width=\linewidth}}
\caption{\small Plots  (a) and (d) show the variation of three Higgs masses 
             with the CP violating phase $\Phi_{\!\mbox{\tiny CP}}$. Plots  (b), (e) and 
             (c) ,(f) show the variation of $ZH_i$ and $WH_i$, $(i =1,2,3)$  production 
            cs at LHC with the phase $\Phi_{\!\mbox{\tiny CP}}$. The left panel  and right 
             panel correspond to the charged Higgs mass of 150 GeV and 200 GeV
             respectively. Other MSSM parameter is fixed to Set~A.} 
\label{fig_3}
\end{figure}
%\newpage
\eject
\begin{figure}[h]
\vspace*{-2.5in}
\centerline{\epsfig{file=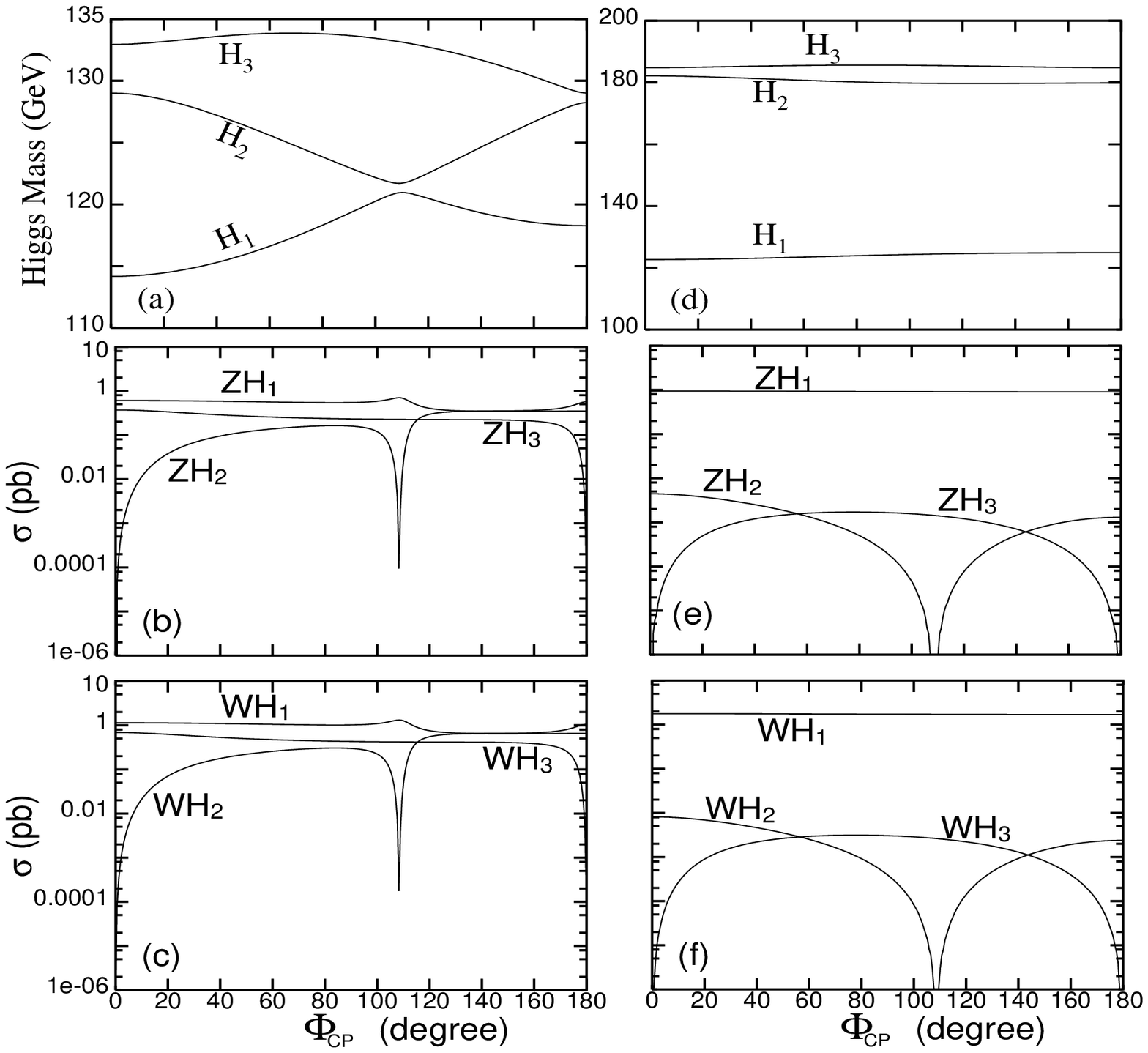,width=\linewidth}}
\caption{\small Plots  (a) and (d) show the variation of three Higgs masses 
             with the CP violating phase $\Phi_{\!\mbox{\tiny CP}}$. Plots  (b), (e) and 
             (c) ,(f) show the variation of $ZH_i$ and $WH_i$, $(i =1,2,3)$ production 
             cs at LHC with the phase $\Phi_{\!\mbox{\tiny CP}}$. The left panel  and right 
             panel correspond to the charged Higgs mass of 150 GeV and 200 GeV
             respectively. Other MSSM parameter is fixed to Set~B.} 
\label{fig_4}
\end{figure}

\end{document}